\relax
\title{{\sc A Re-Evaluation of the\\
nuclear Structure Function Ratios\\
for D, He, $^6\mbox{Li}$, C and Ca}}
\author{}
\date{}
\documentstyle[11pt,epsfig]{article}
\begin{document}
\markright{.}

\maketitle

\begin{center}
THE NEW MUON COLLABORATION (NMC) \\
\vspace{0.8cm}
{\footnotesize{\sl{
Bielefeld~University$^{1+}$,
Freiburg~University$^{2+}$,
Max-Planck~Institut f\"{u}r Kernphysik, Heidelberg$^{3+}$,
Heidelberg~University$^{4+}$,
Mainz~University$^{5+}$, Mons~University$^6$,
Neuch\^{a}tel~University$^7$,
NIKHEF-K$^{8++}$,
Oxford~University$^{9}$,
Saclay DAPNIA/SPP$^{10**}$,
University~of~California, Santa~Cruz$^{11}$,
Paul~Scherrer~Institut$^{12}$,
Torino~University and INFN~Torino$^{13}$,
Uppsala~University$^{14}$,
Soltan~Institute~for~Nuclear~Studies, Warsaw$^{15*}$,
Warsaw~University$^{16*}$,
Wuppertal~University$^{17+}$
}}}\\
\vspace{0.8cm}
{\small{
P.~Amaudruz$^{12v)}$,
M.~Arneodo$^{13a)}$,
A.~Arvidson$^{14}$,
B.~Bade{\l }ek$^{14,16}$,
M.~Ballintijn$^{8c)}$,
G.~Baum$^1$,
J.~Beaufays$^{8b)}$,
I.G.~Bird$^{3,8c)}$,
P.~Bj\"{o}rkholm$^{14}$,
M.~Botje$^{12d)}$,
C.~Broggini$^{7e)}$,
W.~Br\"{u}ckner$^3$,
A.~Br\"{u}ll$^{2f)}$,
W.J.~Burger$^{12g)}$,
J.~Ciborowski$^{8,16}$,
R.~van~Dantzig$^8$,
H.~D\"{o}bbeling$^{3w)}$,
J.~Domingo$^{12x)}$,
J.~Drinkard$^{11)}$,
A.~Dyring$^{14}$,
H.~Engelien$^{2h)}$,
M.I.~Ferrero$^{13}$,
L.~Fluri$^{7\delta)}$,
U.~Gaul$^3$,
P.~Grafstrom$^{14c)}$,
T.~Granier$^{10}$,
D.~von~Harrach$^{3j)}$,
M.~van~der~Heijden$^{8d)}$,
C.~Heusch$^{11}$,
Q.~Ingram$^{12}$,
K.~Janson-Prytz$^{14k)}$,
M.~de~Jong$^{8c)}$,
E.M.~Kabu\ss$^{3j)}$,
R.~Kaiser$^2$,
T.J.~Ketel$^8$,
F.~Klein$^5$,
B.~Korzen$^{17}$,
U.~Kr\"{u}ner$^{17}$,
S.~Kullander$^{14}$,
U.~Landgraf$^2$,
F.~Lettenstr\"{o}m$^{11}$,
T.~Lindqvist$^{14}$,
G.K.~Mallot$^{5}$,
C.~Mariotti$^{13l)}$,
G.~van~Middelkoop$^{8}$,
A.~Milsztajn$^{10}$,
Y.~Mizuno$^{3m)}$,
A.~Most$^3$,
A.~M\"{u}cklich$^3$,
J.~Nassalski$^{15}$,
D.~Nowotny$^{3n)}$,
N.~Pavel$^{17k)}$,
J.~Oberski$^8$,
A.~Pai\'{c}$^7$,
C.~Peroni$^{13}$,
H.~Peschel$^{17y}$,
B.~Povh$^{3,4}$,
R.~Rieger$^{5o)}$,
K.~Rith$^{3p)}$,
K.~R\"{o}hrich$^{5q)}$,
E.~Rondio$^{15}$,
L.~Ropelewski$^{16}$,
A.~Sandacz$^{15}$,
D.~Sanders$^{r)}$,
C.~Scholz$^{3n)}$,
R.~Schumacher$^{12z)}$,
R.~Seitz$^{5u)}$,
U.~Sennhauser$^{12\alpha)}$,
F.~Sever$^{1,8s)}$,
T.-A.~Shibata$^4$,
M.~Siebler$^1$,
A.~Simon$^{3t)}$,
A.~Staiano$^{13}$,
M.~Szleper$^{15}$,
G.~Taylor$^{9\beta)}$,
M.~Treichel$^{3\gamma)}$,
Y.~Tzamouranis$^{3r)}$,
M.~Virchaux$^{10}$,
J.L.~Vuilleumier$^7$,
T.~Walcher$^5$,
R.~Windmolders$^6$,
A.~Witzmann$^2$,
F.~Zetsche$^{3k)}$}} \\

\begin{center}
{\footnotesize\it (accepted by Nuclear Physics)}
\end{center}
\begin{abstract}
\noindent
We present a re-evaluation of
the structure function ratios $F_2^{He}/F_2^{D}$, $F_2^{C}/F_2^{D}$
and $F_2^{Ca}/F_2^{D}$ measured in deep inelastic muon-nucleus  scattering
at an incident muon momentum of 200 GeV. We also present the ratios
$F_2^{C}/F_2^{Li}$, $F_2^{Ca}/F_2^{Li}$ and $F_2^{Ca}/F_2^C$ measured at
90 GeV.
The results are based on data already published
by NMC; the main difference in the analysis is
a correction for the masses of the deuterium targets and an improvement
in the radiative corrections.
The kinematic range covered is $0.0035<x<0.65$, $0.5<Q^2<90$ GeV$^2$
for the He/D, C/D and Ca/D data and $0.0085<x<0.6$, $0.84<Q^2<17$ GeV$^2$
for the Li/C/Ca ones.
\end{abstract}

{\footnotesize {-----------------------------------\\

For footnotes see next page.}}
\newpage

\begin{tabbing}
{}~~~~\=+~~~\=Supported by Bundesministerium f\"{u}r Forschung und
Technologie.\\
\>	++\>	Supported in part by FOM, Vrije Universiteit Amsterdam and NWO.\\
\>	 *\>	Supported by KBN SPUB Nr 621/E - 78/SPUB/P3/209/94. \\
\>	**\>	Supported by CEA, Direction des Sciences de la Mati\a`ere.\\
\> \> \\
\>      a)\>    Now also at Dipartimento di Fisica, Universit\a`a della
Calabria, \\
\>        \>    I-87036 Arcavacata di Rende (Cosenza), Italy. \\
\>	b)\>	Now at Trasys, Brussels, Belgium.\\
\>	c)\>	Now at CERN, 1211 Gen\a`eve 23, Switzerland. \\
\>      d)\>    Now at NIKHEF-H 1009 DB Amsterdam, The Netherlands. \\
\>      e)\>    Now at University of Padova, 35131 Padova, Italy.\\
\>      f)\>    Now at MPI f\"{u}r Kernphysik, 69029 Heidelberg, Germany. \\
\>	g)\>	Now at Universit\a'e de Gen\a`eve, 1211 Gen\a`eve 4, Switzerland.\\
\>	h)\>	Now at LHS GmbH, 63303 Dreieich, Germany.\\
\>      i)\>    Now at University of California, Los Angeles, 90024 Ca, USA.\\
\>      j)\>    Now at University of Mainz, 55099 Mainz, Germany. \\
\>      k)\>    Now at DESY, 22603 Hamburg, Germany.\\
\>	l)\>	Now at INFN-Istituto Superiore di Sanit\a`a, 00161 Roma, Italy.\\
\>	m)\>	Now at Osaka University, 567 Osaka, Japan.\\
\>	n)\>	Now at SAP AG, 69190 Walldorf, Germany.\\
\>      o)\>    Now at Comparex GmbH, 68165 Mannheim, Germany.\\
\>      p)\>    Now at University of Erlangen-N\"{u}rnburg, 91058 Erlangen,
Germany.\\
\>	q)\>	Now at IKP2-KFA, 52428 J\"{u}lich, Germany.\\
\>	r)\>	Now at University of Houston, 77204 Tx, USA.\\
\>      s)\>    Now at ESRF, 38043 Grenoble, France.\\
\>      t)\>    Now at New Mexico State University, Las Cruces Nm, USA.\\
\>      u)\>    Now at Dresden University, 01062 Dresden, Germany.\\
\>      v)\>    Now at TRIUMF, Vancouver, BC V6T 2A3, Canada. \\
\>      w)\>    Now at GSI, 64220 Darmstadt, Germany. \\
\>      x)\>    Now at CEBAF, Newport News, 23606 Va, USA. \\
\>      y)\>    Now at Gruner und Jahr AG \& Co KG, 25524 Itzehoe, Germany. \\
\>      z)\>    Now at Carnegie Mellon University, Pittsburgh, 15213 Pa, USA.
\\
\> $\alpha$)\>    Now at EMPA, 8600 Dubendorf, Switzerland. \\
\> $\beta$)\>    Now at University of Melbourne, Parkville, Victoria 3052,
Australia. \\
\> $\gamma$)\>    Now at Neuch\^{a}tel University, 2000 Neuch\^{a}tel,
Switzerland. \\
\> $\delta$)\>    Now at Bundesamt f\"ur Statistik, 3003 Bern, Switzerland. \\

\end{tabbing}

\vspace{0.8cm}
\end{center}

\newpage
Results on the structure
function ratios $F_2^{He}/F_2^D$, $F_2^{C}/F_2^D$ and
$F_2^{Ca}/F_2^D$~\cite{ourpaper}  as well as $F_2^{C}/F_2^{Li}$,
$F_2^{Ca}/F_2^{Li}$ and $F_2^{Ca}/F_2^C$~\cite{licca}
were recently published by NMC.
The kinematic range covered was $0.0035<x<0.65$, $0.5<Q^2<90$ GeV$^2$
for the He/D, C/D and Ca/D data and $0.0085<x<0.6$, $0.84<Q^2<17$ GeV$^2$
for the $^6\mbox{Li/C/Ca}$ ones.
Here $-Q^2$ is four momentum squared of the virtual photon and
$x=Q^2/(2M\nu)$ is the Bjorken scaling variable, with $M$ the proton
mass and $\nu$ the virtual photon energy in the laboratory frame.
In this paper we present the results of a re-evaluation of
these ratios.

The data were collected using the NMC spectrometer~\cite{f2np} at
the CERN SPS muon beam line at nominal incident energies of
200~GeV for the He/D, C/D and Ca/D data and 90~GeV for
the $^6\mbox{Li/C/Ca}$ ones.

In refs.~\cite{ourpaper,licca} radiative corrections were computed
according to the prescription of Mo and Tsai~\cite{motsai}.
The procedure corrects for the radiative tails of
coherent elastic scattering from nuclei and of quasi-elastic
scattering from nucleons, as well as for  the inelastic radiative
tails. The evaluation of the inelastic tail requires the knowledge
of $F_2$ over a large range of $x$ and $Q^2$. A fit~\cite{f2fit} to the
results of deep inelastic scattering experiments and to
low energy data in the resonance region was used for $F_2^d$.
The bound nucleon structure functions $F_2^A$
were obtained by multiplying $F_2^d$  with  empirical fits to our
cross section ratios together with the SLAC-E139 data~\cite{arnold84}
for $x>0.4$.

Since the publication of refs.~\cite{ourpaper,licca} new measurements of
the structure function $F_2^d$
in the range $0.006<x<0.6$ and $0.5<Q^2<55$ GeV$^2$ have become
available~\cite{nmcf2}. At $x$ less than 0.07
the measured values of $F_2^d$ differ from those of the fit~\cite{f2fit},
by up to 18\% at $x=0.0035$ and $Q^2=0.6$ GeV$^2$.
Furthermore the SLAC structure function ratios
have been reanalysed~\cite{reanalysis}. These new ratios
differ from the old ones by up to a few per cent.

In addition, it was found that the masses of the liquid deuterium targets
used in ref.~\cite{ourpaper} for the
C/D and Ca/D data had been incorrectly evaluated.
Correcting these has increased the corresponding structure function ratios
by 0.74\%.

We therefore recomputed the radiative corrections using
the new $F_2^d$ data, the results of
the SLAC reanalysis  and the correct deuterium target
masses for our C/D and Ca/D ratios.
This has resulted in an increase of the structure function ratios
at the smallest values of $x$ which is negligible for the He/D data but
ranges up to 2.5\% for Ca/D.

Radiative corrections were calculated with three different programs.
The first was the one used to obtain the results presented
in refs.~\cite{ourpaper,licca}, based on
the Mo and Tsai formalism; the second was an improved version of the first
including vacuum polarisation by quark and $\tau$ loops and electroweak
interference terms. The third program is based on the covariant
approach described in ref.~\cite{dubna}. The results obtained with the three
methods
are consistent. The ratios presented in this paper were obtained with
the third method.

\begin{figure}[p]
\begin{centering}
\epsfig{figure=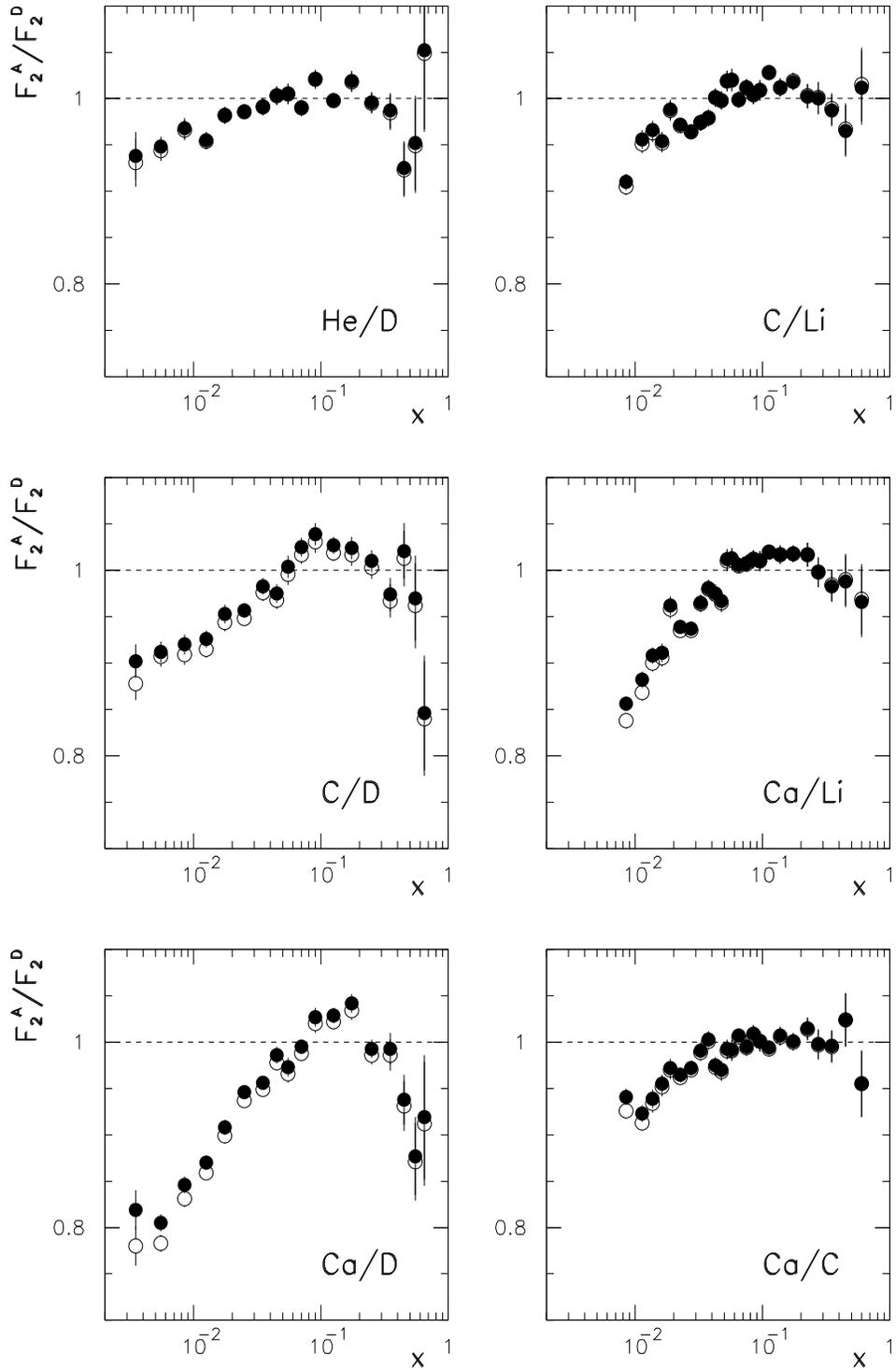,height=8.5in,width=5.5in}
\caption{Structure function ratios as function of $x$, averaged over $Q^2$.
The full circles represent the re-evaluated ratios, the open circles the
old ratios. Only statistical errors are shown.}
\label{fig:fig1}
\end{centering}
\end{figure}

\begin{figure}[t]
\begin{centering}
\epsfig{figure=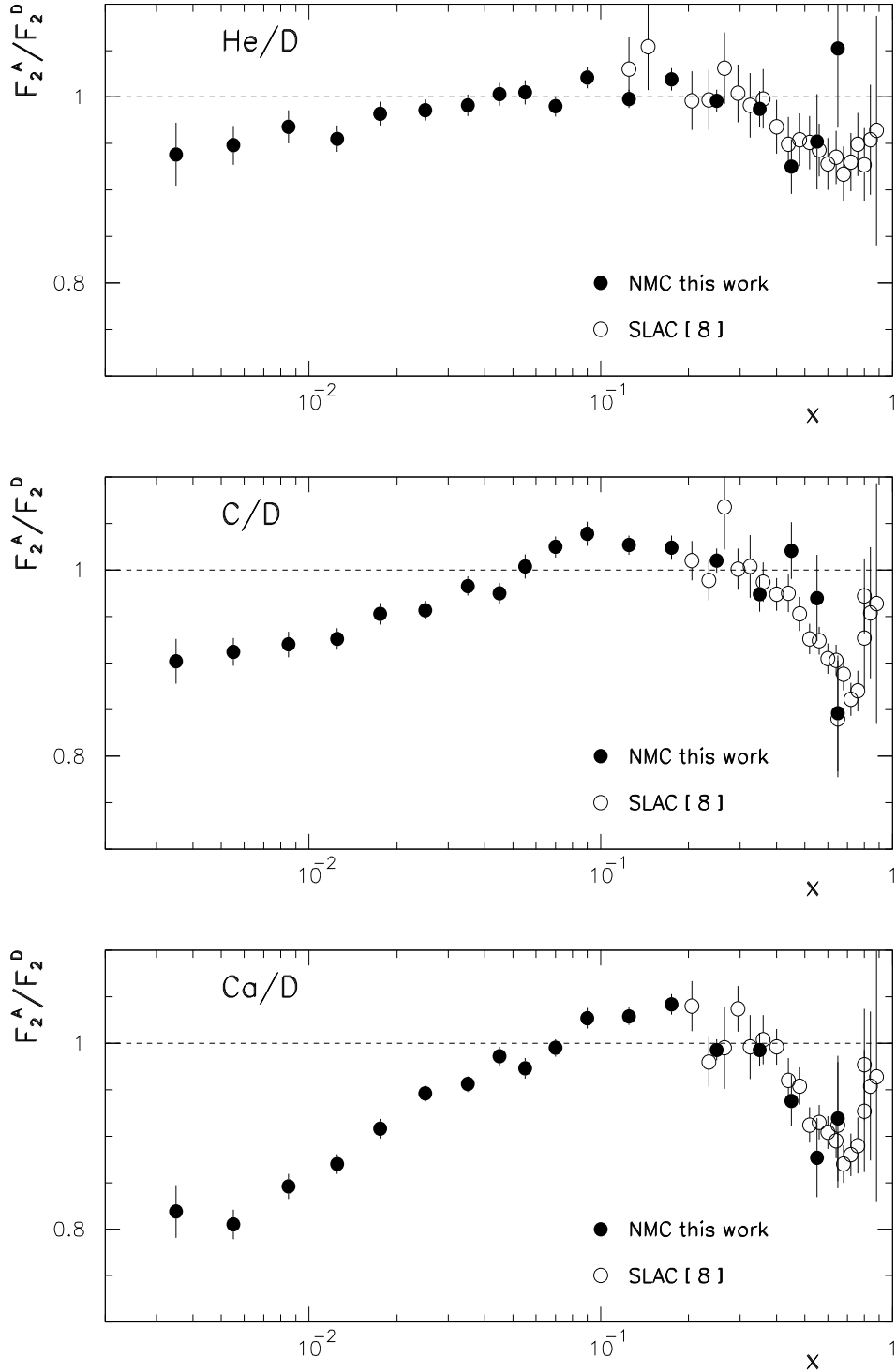,height=8.5in,width=5.5in}
\caption{The re-evaluated NMC structure function ratios
for He/D,~C/D,~Ca/D together with the reanalysed SLAC results.
The error bars show the statistical and systematic errors added in
quadrature. The normalisation uncertainties are not included.}
\label{fig:fig2}
\end{centering}
\end{figure}

\begin{figure}[t]
\begin{centering}
\epsfig{figure=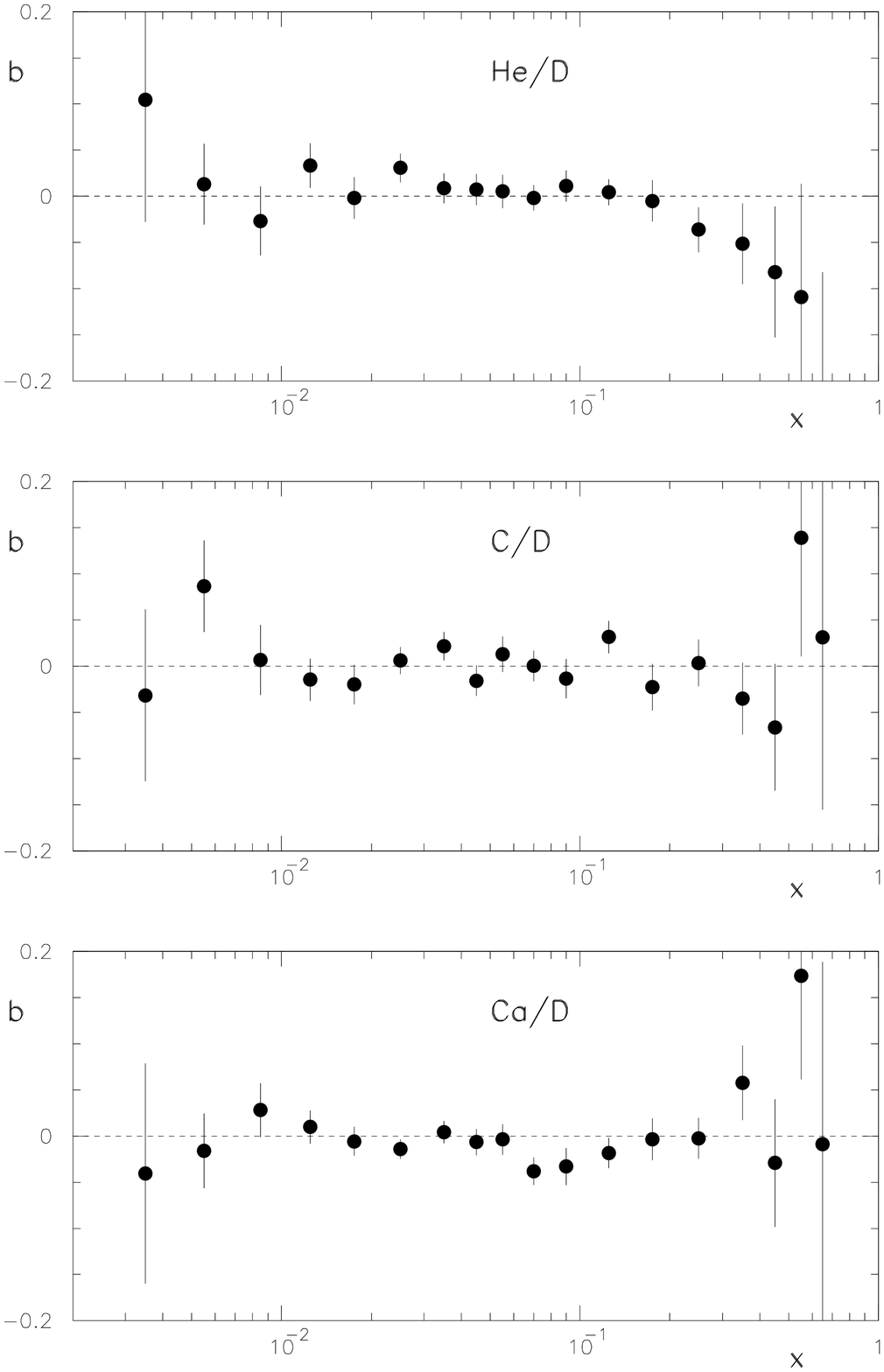,height=8.5in,width=5.5in}
\caption{The slopes $b$ from a linear fit in $\ln Q^2$ for each $x$ bin
separately. The errors shown are statistical only.}
\label{fig:fig3}
\end{centering}
\end{figure}

In fig. \ref{fig:fig1}
the present results are shown as a function of $x$ and
are compared with the old ones. The new results are also given in
table \ref{tab:rat200} and table \ref{tab:rat090} for the measurements at
200 GeV and 90 GeV, respectively.
In fig. \ref{fig:fig2}
we present our data for He/D, C/D and Ca/D
together with the reanalysed SLAC
results~\cite{reanalysis}.
Finally, fig. \ref{fig:fig3} shows the
logarithmic $Q^2$ slopes $b$ obtained from fits of the form
$F_2^{A}/F_2^{D}\! = \!a\! +\! b\ln {Q^2}$ to the He/D, C/D and Ca/D
ratios in each $x$ bin separately. The $Q^2$ dependences are essentially
unchanged with respect to those presented in ref.~\cite{ourpaper}.

\begin{table}[t]
\vspace*{-2.0cm}
\begin{centering}
\footnotesize
\begin{tabular}{cccccc}
\hline \\
$ x $ & $\langle Q^2 \rangle$ &
$\langle y \rangle$ &
$F_2^{He}/F_2^{D}$ & stat & syst \\
  &  [$\mbox{GeV}^2$] & & & & \\
\hline
0.0035&  0.77&   0.65& 0.938& 0.026& 0.022  \\
0.0055&   1.3&   0.65& 0.948& 0.011& 0.018  \\
0.0085&   1.8&   0.59& 0.968& 0.011& 0.014  \\
0.0125&   2.4&   0.53& 0.955& 0.009& 0.011  \\
0.0175&   3.0&   0.47& 0.982& 0.009& 0.009  \\
0.025 &   3.8&   0.42& 0.986& 0.008& 0.008  \\
0.035 &   4.7&   0.37& 0.991& 0.009& 0.007  \\
0.045 &   5.6&   0.34& 1.003& 0.010& 0.007  \\
0.055 &   6.3&   0.31& 1.005& 0.011& 0.007  \\
0.070 &   7.3&   0.29& 0.990& 0.009& 0.006  \\
0.090 &   8.7&   0.26& 1.021& 0.010& 0.006  \\
0.125 &  11  &   0.24& 0.998& 0.008& 0.006  \\
0.175 &  14  &   0.22& 1.019& 0.011& 0.005  \\
0.25  &  19  &   0.21& 0.996& 0.011& 0.005  \\
0.35  &  24  &   0.19& 0.987& 0.019& 0.005  \\
0.45  &  31  &   0.19& 0.925& 0.029& 0.005  \\
0.55  &  38  &   0.19& 0.952& 0.051& 0.005  \\
0.65  &  44  &   0.18& 1.052& 0.085& 0.005  \\
\hline \\
$ x $ & $\langle Q^2 \rangle$ &
$\langle y \rangle$ &
$F_2^{C}/F_2^{D}$ & stat & syst \\
  &  [$\mbox{GeV}^2$] & & & & \\
\hline
0.0035&  0.74&    0.62& 0.902& 0.018& 0.016  \\
0.0055&   1.2&    0.57& 0.912& 0.011& 0.010  \\
0.0085&   1.7&    0.54& 0.920& 0.011& 0.008  \\
0.0125&   2.3&    0.51& 0.926& 0.009& 0.007  \\
0.0175&   3.0&    0.47& 0.953& 0.010& 0.006  \\
0.025 &   3.8&    0.42& 0.957& 0.008& 0.006  \\
0.035 &   4.9&    0.38& 0.983& 0.009& 0.005  \\
0.045 &   6.0&    0.37& 0.975& 0.010& 0.005  \\
0.055 &   7.2&    0.36& 1.004& 0.012& 0.005  \\
0.070 &   8.8&    0.34& 1.025& 0.010& 0.005  \\
0.090 &  11  &    0.33& 1.039& 0.012& 0.005  \\
0.125 &  14  &    0.31& 1.027& 0.009& 0.005  \\
0.175 &  17  &    0.27& 1.024& 0.012& 0.005  \\
0.25  &  21  &    0.23& 1.010& 0.012& 0.005  \\
0.35  &  26  &    0.21& 0.974& 0.018& 0.005  \\
0.45  &  31  &    0.19& 1.021& 0.030& 0.006  \\
0.55  &  37  &    0.19& 0.970& 0.046& 0.006  \\
0.65  &  42  &    0.17& 0.846& 0.062& 0.007  \\
\hline \\
$ x $ & $\langle Q^2 \rangle$ &
$\langle y \rangle$ &
$F_2^{Ca}/F_2^{D}$ & stat & syst \\
  &  [$\mbox{GeV}^2$] & & & & \\
\hline
0.0035&  0.60&    0.47& 0.819& 0.021& 0.019  \\
0.0055&  0.94&    0.46& 0.805& 0.009& 0.013  \\
0.0085&   1.4&    0.43& 0.846& 0.009& 0.010  \\
0.0125&   1.9&    0.41& 0.870& 0.007& 0.008  \\
0.0175&   2.5&    0.40& 0.908& 0.008& 0.007  \\
0.025 &   3.4&    0.38& 0.946& 0.006& 0.006  \\
0.035 &   4.7&    0.36& 0.956& 0.007& 0.005  \\
0.045 &   5.7&    0.35& 0.986& 0.009& 0.005  \\
0.055 &   6.8&    0.34& 0.973& 0.010& 0.005  \\
0.070 &   8.1&    0.32& 0.995& 0.008& 0.005  \\
0.090 &   9.7&    0.30& 1.027& 0.010& 0.005  \\
0.125 &  12  &    0.26& 1.029& 0.008& 0.005  \\
0.175 &  14  &    0.23& 1.042& 0.010& 0.005  \\
0.25  &  19  &    0.21& 0.993& 0.010& 0.005  \\
0.35  &  24  &    0.19& 0.993& 0.017& 0.006  \\
0.45  &  30  &    0.19& 0.938& 0.027& 0.006  \\
0.55  &  35  &    0.18& 0.877& 0.042& 0.006  \\
0.65  &  41  &    0.17& 0.919& 0.067& 0.007  \\
\hline\end{tabular}
\caption{The structure function ratios $F_2^A/F_2^D$ measured at 200 GeV and
averaged over $Q^2$.
The normalisation uncertainty of 0.4\% is not included in the systematic
errors.
The variable $y$ is defined as $\nu /E$, where $E$ is the incident muon
energy.}
\label{tab:rat200}
\end{centering}
\end{table}

\begin{table}[t]
\vspace*{-2.0cm}
\begin{centering}
\footnotesize
\begin{tabular}{ccc|ccc|ccc|ccc}
\hline \\[-0.375cm]
& & & & & & & & & & & \\
$ x $ & $\langle Q^2 \rangle$ &
$\langle y \rangle$ &
$F_2^{C}/F_2^{Li}$ & stat & syst &
$F_2^{Ca}/F_2^{Li}$ & stat & syst &
$F_2^{Ca}/F_2^{C}$ & stat & syst \\
&  [$\mbox{GeV}^2$] & & & & & & & & & & \\
\hline
0.0085&     0.8& 0.57& 0.910& 0.009& 0.012& 0.856& 0.008& 0.022& 0.941& 0.009&
0.023  \\
0.0113&     1.1& 0.59& 0.956& 0.010& 0.009& 0.882& 0.009& 0.017& 0.923& 0.009&
0.015  \\
0.0138&     1.2& 0.53& 0.966& 0.010& 0.007& 0.908& 0.009& 0.013& 0.939& 0.010&
0.011  \\
0.0163&     1.4& 0.52& 0.954& 0.010& 0.005& 0.911& 0.010& 0.010& 0.955& 0.010&
0.008  \\
0.0188&     1.6& 0.52& 0.988& 0.011& 0.004& 0.962& 0.010& 0.008& 0.972& 0.010&
0.007  \\
0.0225&     1.8& 0.49& 0.972& 0.008& 0.004& 0.939& 0.007& 0.006& 0.965& 0.008&
0.005  \\
0.0275&     2.0& 0.44& 0.964& 0.008& 0.003& 0.937& 0.008& 0.004& 0.972& 0.008&
0.003  \\
0.0325&     2.2& 0.41& 0.974& 0.009& 0.002& 0.965& 0.009& 0.003& 0.991& 0.009&
0.003  \\
0.0375&     2.3& 0.37& 0.979& 0.009& 0.002& 0.981& 0.009& 0.002& 1.003& 0.009&
0.002  \\
0.0425&     2.4& 0.34& 1.001& 0.010& 0.002& 0.975& 0.010& 0.002& 0.975& 0.009&
0.002  \\
0.0475&     2.6& 0.33& 0.998& 0.010& 0.002& 0.967& 0.010& 0.002& 0.970& 0.010&
0.002  \\
0.0525&     2.7& 0.31& 1.019& 0.011& 0.002& 1.012& 0.011& 0.002& 0.993& 0.010&
0.002  \\
0.0575&     2.8& 0.29& 1.020& 0.012& 0.002& 1.013& 0.011& 0.002& 0.992& 0.011&
0.002  \\
0.065 &     3.0& 0.28& 0.999& 0.009& 0.002& 1.006& 0.008& 0.002& 1.007& 0.008&
0.002  \\
0.075 &     3.3& 0.27& 1.012& 0.009& 0.002& 1.008& 0.009& 0.002& 0.995& 0.009&
0.002  \\
0.085 &     3.6& 0.26& 1.004& 0.010& 0.002& 1.012& 0.010& 0.002& 1.009& 0.010&
0.002  \\
0.095 &     3.9& 0.25& 1.009& 0.011& 0.002& 1.010& 0.011& 0.002& 1.001& 0.011&
0.002  \\
0.113 &     4.3& 0.23& 1.028& 0.008& 0.002& 1.020& 0.008& 0.002& 0.994& 0.008&
0.002  \\
0.138 &     5.1& 0.22& 1.012& 0.010& 0.002& 1.017& 0.010& 0.002& 1.007& 0.010&
0.002  \\
0.175 &     6.2& 0.22& 1.018& 0.009& 0.002& 1.018& 0.009& 0.002& 1.001& 0.009&
0.002  \\
0.225 &     7.7& 0.21& 1.002& 0.013& 0.002& 1.017& 0.013& 0.002& 1.015& 0.012&
0.002  \\
0.275 &     9.1& 0.20& 1.000& 0.017& 0.002& 0.998& 0.016& 0.002& 0.998& 0.016&
0.002  \\
0.35  &    11  & 0.19& 0.987& 0.017& 0.002& 0.983& 0.017& 0.002& 0.996& 0.017&
0.002  \\
0.45  &    14  & 0.18& 0.965& 0.028& 0.002& 0.988& 0.028& 0.002& 1.024& 0.029&
0.002  \\
0.60  &    17  & 0.17& 1.012& 0.040& 0.004& 0.966& 0.038& 0.004& 0.955& 0.036&
0.002  \\
\hline\end{tabular}
\caption{The nuclear structure function ratios measured at 90 GeV and averaged
over $Q^2$.
The normalisation uncertainties (not included in the systematic errors)
are 0.7\%, 0.8\% and 0.5\% for C/Li, Ca/Li and Ca/C, respectively.
The variable $y$ is defined as $\nu /E$, where $E$ is the incident muon
energy.}
\label{tab:rat090}
\end{centering}
\end{table}

\newpage
\begin{thebibliography}{99}
\bibitem{ourpaper} CERN NA37/NMC, P. Amaudruz et al., Z. Phys.{\bf~C 51}
(1991) 387
\bibitem{licca} CERN NA37/NMC, P. Amaudruz et al., Z. Phys.{\bf ~C 53} (1992)
73
\bibitem{f2np} CERN NA37/NMC, P. Amaudruz et al., Nucl. Phys.{\bf ~B371} (1992)
3
\bibitem{motsai}L.W. Mo and Y.S. Tsai, Rev. Mod. Phys. {\bf 41} (1969) 205;\\
      { Y. S. Tsai, SLAC-PUB-848 (1971)}
\bibitem{f2fit} CERN NA37/NMC, P. Amaudruz et al., Phys. Rev. Lett. 66 (1991)
2712
\bibitem{arnold84} SLAC-E139, R.G. Arnold et al.,
     Phys. Rev. Lett. {\bf 52} (1984) 727;\\
 R.G. Arnold, SLAC-PUB-3257 (1983)
\bibitem{nmcf2} CERN NA37/NMC, P. Amaudruz et al.,
Phys. Lett. {\bf B 295} (1992) 159
\bibitem{reanalysis} SLAC-E139, J. Gomez et al., Phys. Rev.{\bf ~D49} (1994)
4348
\bibitem{dubna}A. Akhundov et al., DESY 94-115 (1994);
CERN-TH 7339/94 (1994); IC/94/154 (1994)\\
\end{thebibliography}
\end{document}